# The effect of dispersal area on the extinction threshold

Róbert Juhász[1], Igor D. Kovács[2] and Beáta Oborny[2,3]

[1]HUN-REN Wigner Research Centre for Physics, Institute for Solid State Physics and Optics,

H-1525 Budapest, P.O. Box 49, Hungary.

RJ: juhasz.robert@wigner.hun-ren.hu

[2]Institute of Biology, Eötvös Loránd University,

H-1117 Budapest, Pázmány P. stny. 1C, Hungary.

BO: beata.oborny@ttk.elte.hu; IDK: igorjee@gmail.com

[3]Institute of Evolution, HUN-REN Centre for Ecological Research,

H-1121 Budapest, Konkoly-Thege M. Road 29-33, Hungary

BO: oborny.beata@ecolres.hu

Corresponding author: Beáta Oborny (beata.oborny@ttk.elte.hu)

Abstract

The survival of populations hinges on their ability to offset local extinctions through new colonizations. The dispersal area ($A$) plays a crucial role in this process, as it determines the probability of finding colonizable vacant sites. We investigated the spatial colonization-extinction dynamics in a lattice model (a contact process), exploring various finite dispersal areas ($A$) and estimating the extinction threshold $\lambda_E(A)$. Our results revealed a consistent $\lambda_E(A)$ relationship, largely independent of lattice geometry (except for the smallest $A$). This $\lambda_E(A)$ relationship obeyed universal scaling laws within two broad ranges of $A$. The scaling relations suggest considerable selection upon the increase of dispersal area, particularly at low $A$ values. We discuss these findings in the broader context of the evolution of dispersal area.

## 1. Introduction

The extinction threshold is a fundamental concept in population ecology, marking the point below which populations cannot sustain themselves demographically, as local extinctions cannot be counterbalanced by new colonizations (Levins, 1969; Holt and Keitt, 2000; Oborny et al., 2005; Ovaskainen et al., 2020). Understanding and identifying these thresholds is key to predicting the long-term persistence of a species and informing conservation strategies to reduce extinction risk (Akcakaya and Sjögren-Gulve, 2000; Oborny et al., 2007).

The area over which individuals (e.g., seeds) are dispersed can greatly influence the success of colonization, in at least two important ways. 1) It affects clumping, and thus, intraspecific competition (Ovaskainen et al., 2020; Law et al., 2003). 2) It affects the likelihood of reaching new habitat patches in case of habitat loss and fragmentation (Hanski, 1998; Baguette et al., 2013). The first effect arises even in homogeneous environments, while the second assumes habitat heterogeneity.

Several models have sought to explore the interplay between demographic and dispersal traits in determining metapopulation persistence. (e.g., Travis et al., 2012; de Oliviera et al. 2020; Gattringer et al., 2023; see reviews by Ronce, 2007; Bonte et al., 2012; Duputié and Massol, 2013). Most of these studies assumed patchy environments, typically by interspersing habitable sites with uninhabitable ones ("diluting a lattice"; e.g., Bascompte and Solé, 1996; Keymer et al., 2000; King and With, 2002; Finand et al., 2023), or assuming that habitable sites are nodes within a network (e.g., Hanski 1994; Hanski and Ovaskainen 2000). Thus, the aforementioned effects 1 and 2 were combined. These studies often highlight the intrinsic complexity of effect 2 arising from the variety of possible spatial configurations of the habitable patches (Stauffer and Aharony 1992). For example, diluted lattice models have identified a critical threshold in the connectivity of habitable patches (the percolation threshold) which can fundamentally influence metapopulation persistence (Oborny et al., 2007). Because of the inherent complexity of effect 2, effect 1 is still poorly understood. How does limited dispersal impact the clumping of occupied sites, and in turn, how does this influence the extinction threshold? Our goal is to address these questions within homogeneous environments, excluding effect 2.

For this purpose, we use a lattice model in which all lattice sites are uniformly suitable, characterized by a constant colonization rate ($c$) and local extinction rate ($e$). We vary $c$ and $e$ and estimate the extinction threshold, i.e., the ratio of the demographic parameters $\lambda = \frac{c}{e}$ at which the steady-state occupancy becomes zero. We investigate the effect of the dispersal area ($A$) on the extinction threshold, $\lambda_E(A)$.

Each lattice cell can host either a single individual or multiple individuals, depending on the model's interpretation. In the former case, the model is interpreted as a spatial logistic model of a population with stochastic birth and death; whereas in the latter, it corresponds to a spatial version of the Levins metapopulation model with stochastic colonization and extinction. We adopt the colonization-extinction interpretation throughout the text and revisit both interpretations in the Discussion.

We hypothesize the existence of a universal scaling law that captures $\lambda_E(A)$ independently of the underlying lattice geometry, and across a broad range of $A$. This hypothesis is motivated by earlier studies showing that extinction in dispersal-limited populations can be viewed as a phase transition from the living to the extinct state, governed by well-defined scaling laws (Harris 1974; Liggett 1999; Oborny et al. 2007; Solé 2011). First, we briefly review the literature on the extinction threshold, focusing on the effect of $\lambda$ on the spatial behavior of populations near the threshold. Second, we revisit some results from probability theory and spatial stochastic modelling which predict the impact of dispersal area on the extinction threshold in the limit of infinitely large dispersal area. Third, we apply spatially explicit computer simulations to explore this relationship for more realistic, small dispersal areas. Finally, we discuss the results in the context of the evolution of dispersal area.

The main novelty of our research is evaluating the extinction threshold at a broad diversity of dispersal areas, ranging from the nearest "contact" dispersal to a well-mixed system with unlimited dispersal. We believe that the inclusion of short-distance dispersal (small *A* values) into the study of $\lambda_E(A)$ represents a considerable step toward modeling real species. We discuss the plausible ranges of *A* in plant populations and metapopulations (**Section 4.3**). In general, universality of the scaling relations motivates the search for common patterns in the extinction processes of different species across diverse environments.

## 2. Methods

### *2.1. The spatial population dynamic model*

Our model is based on the contact process (CP; Harris 1974), a foundational framework for studying population dynamics with neighborhood interactions (see the 'Limited Dispersal' section for details). We extend the original CP by exploring various neighborhood sizes and lattice geometries. Specifically, we examine a square lattice with three different types of dispersal geometries and a hexagonal lattice. In each of these cases, we analyze a series of dispersal neighborhoods of varying sizes (**Figure 1**; see **Supplement 1** for a general, formal definition of neighborhood). The dispersal area *A* is defined as the number of cells belonging to the neighborhood. We use *A* rather than the dispersal radius to make the results obtained at different lattice geometries comparable. Viewing the lattice as a regular network of sites (cells), *A* is consistently the number of links of a cell along which dispersal can occur (i.e., its degree). The lattice sizes range from $2.5 \cdot 10^4$ to $4 \cdot 10^{12}$ cells.

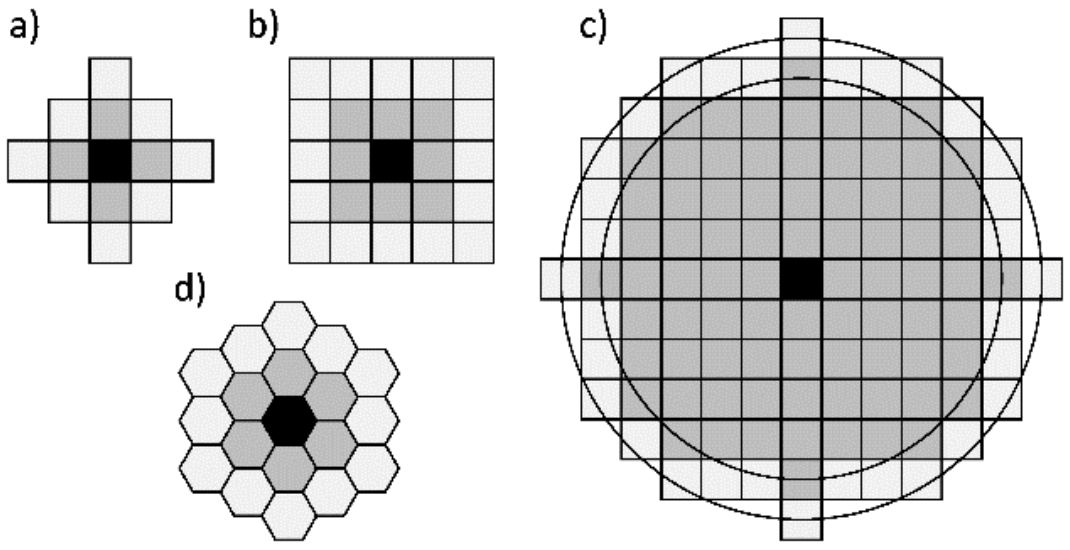

**Figure 1.** Illustration of different neighbourhoods in our simulations. Each sub-figure shows an example for a smaller (dark grey) and a larger (light grey) neighborhood around the focal cell (black). The dispersal area (*A*) is defined as the number of cells within the neighborhood. **a)** Diamond, *A*= 4 or 12. **b)** Square, *A*= 8 or 24. **c)** Circle, *A*= 80 or 112. At small *A* values, the circle and the diamond are equivalent. **d)** Hexagon, *A*= 6 or 18.

Having defined the neighborhood of a given lattice cell, we can now specify the model as follows. Each cell can be in either of two states: empty or occupied. The dynamics of occupancy are modeled by a stochastic cellular automaton. In each elementary updating step, we select a cell randomly from the set of occupied cells. Then it becomes vacant with probability $e/(c+e)$, or attempts to colonize another cell with probability $c/(c+e)$ (following a standard method in stochastic Markov processes; see page 162 in Marro and Dickman 1999). This target cell is selected randomly, with equal probabilities from the source cell's neighborhood (**Figure 1**). The colonization is successful if the selected target cell is empty. The elementary step is repeated many times. Time is measured in Monte Carlo Steps (MCS). Within an MCS, each cell is updated once on average. Therefore, the updating of an occupied cell increases time by $1/N$, where $N$ is the number of occupied cells. This updating method is efficient in terms of computer time, especially near the extinction threshold, where the proportion of occupied sites is low.

It is important to note that $c$ is merely an attempt rate of colonization. The success depends on whether the target cell is vacant or not. Since the occupied sites are clumped in the case of limited dispersal, the dispersal area ($A$) also influences the success of colonization (**Figure 2**).

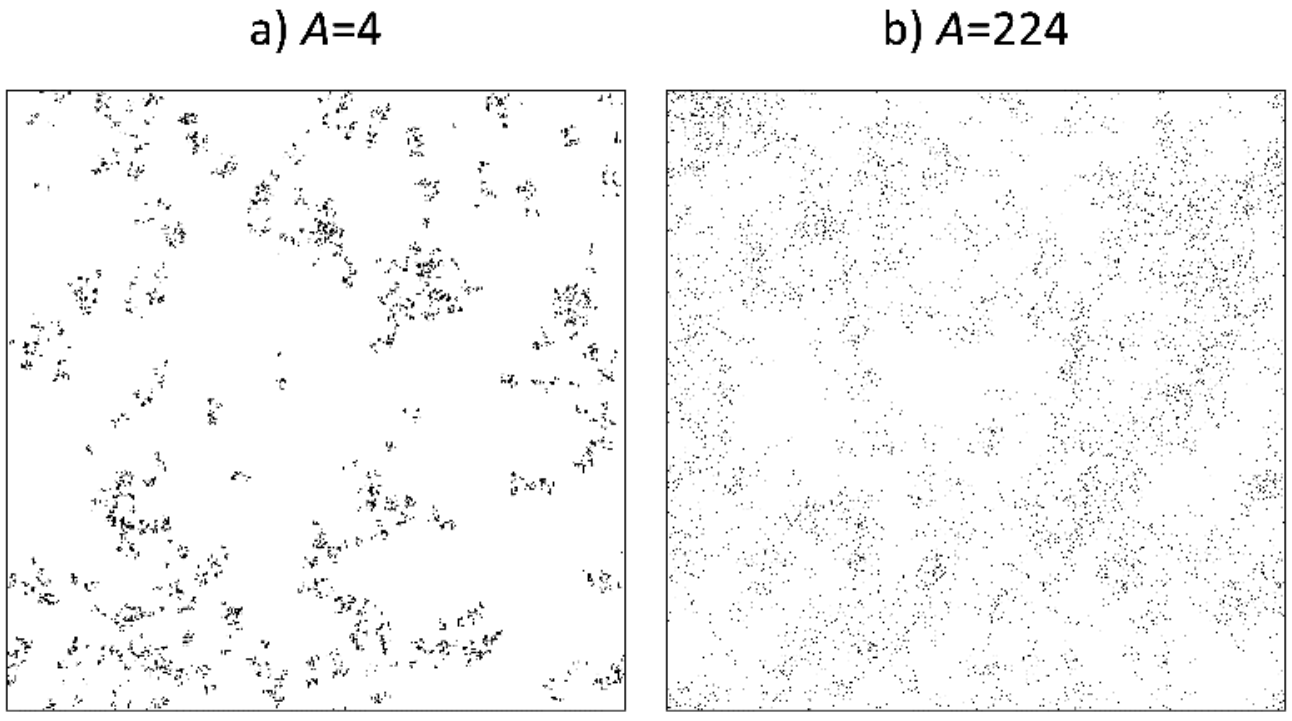

**Figure 2.** Examples of steady-state occupancy patterns in the case of different dispersal areas ($A$). The $\lambda$ values (**a:** 1.65; **b:** 1.05) were selected to present similar total densities (**a:** 0.0239; **b:** 0.0231). These simulations were conducted on a 500 × 500 lattice, that is smaller than the lattice sizes used to obtain the results (see **Table 1**).

### *2.2. Theoretical framework for the analysis of model output*

The analysis of our results is based on some earlier foundational results in theoretical ecology and statistical physics. First, we summarize key findings from the Levins model (Levins 1969), which assumes unlimited dispersal. Therefore, it serves as an important baseline for our analysis (a mean field approximation). Secondly, we present the classical contact process (Harris 1974), a spatially explicit model, in which the characteristic phase transition at the extinction threshold was first described. This model assumes strict dispersal limitation, with $A$=4. We then relax this strict assumption allowing $4 \leq A < \infty$, and present our simulations.

#### 2.2.1. Unlimited dispersal

In a lattice model with unlimited dispersal, colonization is attempted with equal probability in every cell. In the limit of infinite area, this is equivalent to the well-known Levins model of patch occupancy (Levins 1969),

$$\frac{dn}{dt} = cn(1-n) - en, \qquad \text{Eq. 1}$$

where *n* is the proportion of occupied sites. (*1-n*) expresses that the vacant sites are equally available. The expected duration of occupancy (lifetime) in each cell is *1/e*. A frequently used control parameter in this system is $\lambda = \frac{c}{e}$, the 'demographic rate' (Holt et al., 2005). The Levins model of metapopulations is formally equivalent to the logistic model of ordinary populations. The steady-state population size is

$$\hat{n} = 1 - \frac{1}{\lambda}. \qquad \text{Eq. 2}$$

In the extinct state, $\hat{n} = 0$. The extinction threshold in this system is at $\lambda$=1 (see the dashed line in **Figure 3**). We will use this value as a reference: the extinction threshold in case of unlimited dispersal, denoted by $\lambda_E(\infty)$.

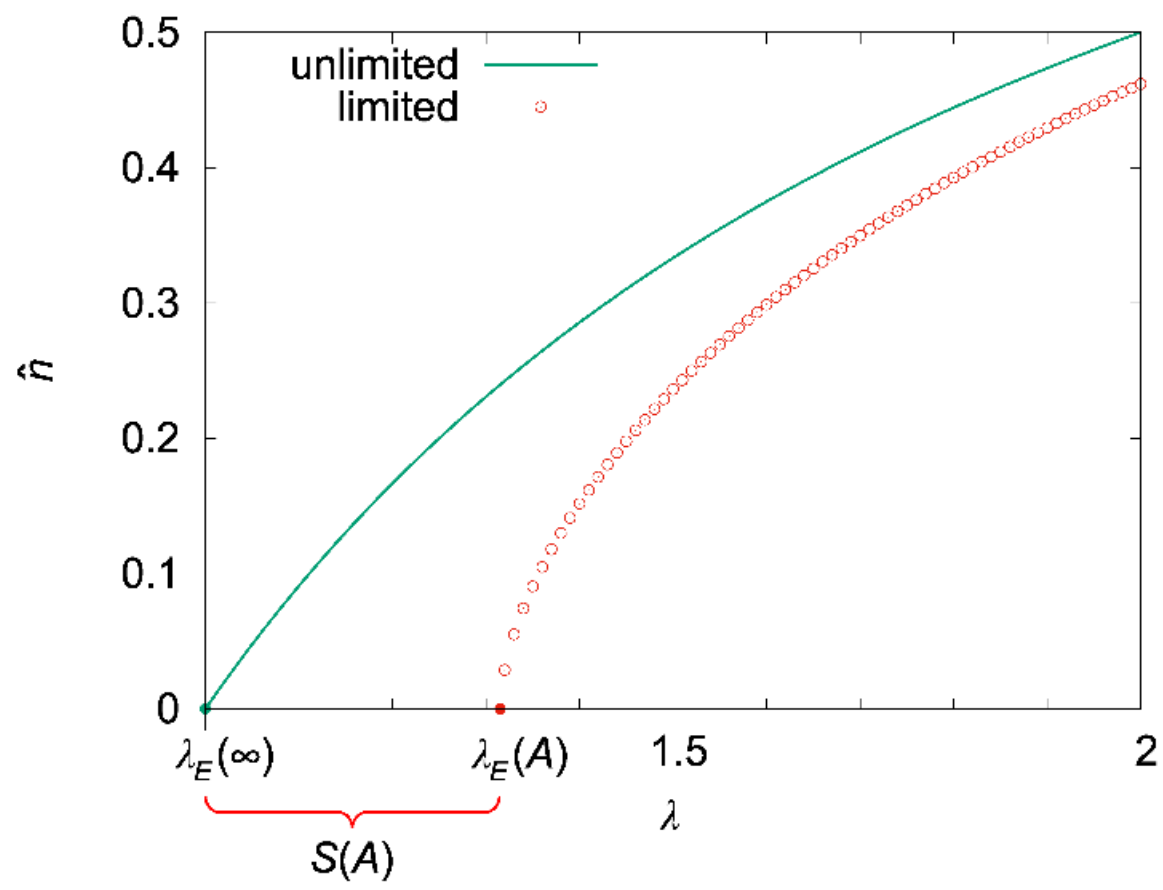

**Figure 3.** Dependence of the steady-state occupancy ($\hat{n}$) on the demographic rate ($\lambda = c/e$). The continuous curve shows the case of unlimited dispersal (Equation 2), while the dots have been obtained from our simulations in the case of limited dispersal (using a circular dispersal area with *A*=12). The extinction threshold is $\lambda_E(\infty) = 1$ in the unlimited case, while it shifts to $\lambda_E(A) > \lambda_E(\infty)$ under dispersal limitation, depending on the dispersal area (*A*). The shift is *S*(*A*).

2.2.2. Limited dispersal

A spatially explicit version of the Levins model with limited dispersal was first introduced in probability theory (Harris 1974) and has been thoroughly studied in statistical physics (Liggett 1999; Marro and Dickman 1999; Ódor 2004; Henkel et al., 2008) under the name 'contact process' (CP). The CP gradually became a broadly used model in ecology as a simple spatial version of the Levins metapopulation (or logistic population) model (Oborny et al., 2007; Solé 2011; Ovaskainen et al., 2020). The original CP applied diamond lattice geometry with $A$=4, a convention that was followed by almost all subsequent studies. These theoretical investigations revealed that extinction in the CP is a critical phase transition at a sharp threshold $\lambda_E(A)$ (see the solid line in **Figure 3**), where $A$ denotes a finite dispersal area.

The CP shows several exciting features in the limit of infinite system size. The steady-state occupancy changes according to a characteristic scaling law in the vicinity of the threshold (in the critical region):

$$\hat{n} \propto [\lambda - \lambda_E(A)]^{\beta}, \qquad \text{Eq. 3}$$

where $\propto$ stands for 'proportional to'. Monte Carlo simulations have estimated $\beta$=0.584(4) in two dimensions (Marro and Dickman 1999; Henkel et al., 2008). The number in brackets shows the error of the estimate. Interestingly, the value of $\beta$ does not depend on the lattice geometry or neighborhood size, only on the dimensionality. This kind of universality is common for critical phase transitions. A heuristic explanation for this universal feature is that the spatial correlation length diverges at the transition point, therefore the cell-scale details of the lattice become irrelevant. Due to limited dispersal, the spatial pattern of occupied sites is clumped (**Figure 2**), i.e., the local density in the neighborhood of an occupied cell is typically higher than the global density. Consequently, self-regulation becomes deficient. This can directly be detected by a peak of the spatial and temporal variance of occupancy within finite areas (Harris 1974; Stanley 1987), which may serve as an early warning sign of extinction (Oborny et al., 2005). It is noteworthy that the scaling laws that govern the extinction process apply to a whole universality class of (meta)population models, the so-called directed percolation universality class, a representative of which is the CP (Broadbent and Hammersley 1957).

The extinction threshold is non-universal, as its value does depend on the lattice geometry and dispersal area. This may explain why $\lambda_E(A)$ has been out of the focus of most studies. To our knowledge, only a small number of papers have attempted to derive a mathematical formula for $\lambda_E(A)$, based on different methods of approximation (Bramson et al., 1989; Durrett and Perkins 1999; Ovaskainen et al., 2020). We summarize these studies in **Section 3** in comparison with our results. We evaluate the applicability of these asymptotic forms in a broad range of *A*, including very small *A* values (local dispersal), and estimate the scale of the dispersal area within which the asymptotic form can be used with good accuracy. We aim to facilitate the applicability of these earlier results to real plant populations.

**Table 1** summarizes the lattice sizes and the ranges of *A* values used in the simulations. *A* was increased stepwise from the minimum to the maximum by adding a new layer of available cells in each step (like in **Figure 1**), except for circle-shaped neighborhoods, where the dispersal radii were integer powers of 2.

| Lattice geometry | Lattice size (cells) | Range of *A* (cells) |
|---|---|---|
| Diamond | $2.5{\cdot}10^4$-$4{\cdot}10^6$ | 4 – 180 |
| Hexagon | $2.5{\cdot}10^4$-$4{\cdot}10^6$ | 6 – 216 |
| Square | $2.5{\cdot}10^4$-$4{\cdot}10^6$ | 8 – 440 |
| Circle | $4{\cdot}10^{12}$ | 12 –51,432 |

**Table 1.** Lattice areas and ranges of dispersal areas (*A*) in the numerical simulations.

*2.3. Estimation of the extinction threshold*

We estimated $\lambda_E(A)$ for finite *A* using numerical simulations. Each run started from a single occupied site at the center of the lattice; the other sites were vacant. The simulations run in $10^4$–$10^6$ independent repetitions at every parameter combination using the contact process rules described above. The repetitions were stopped simultaneously, at the latest when any of them reached the edge of the lattice. Thus, the effects of finite lattice size were precluded. We estimated the extinction threshold by studying the dynamics of spreading. Three characteristics were recorded in each set of runs: the survival probability, the mean number of individuals, and the mean squared displacement from the origin (see the **Appendix**, where the methods are described in detail). Plotting these quantities over time (*t*) on a log-log plot, each tended to a straight line at large *t* values as the extinction threshold is approached [$\lambda \to \lambda_E(A)$]. Changing $\lambda$ in small steps, we approximated the extinction threshold to four digits.

## 3. Results

### *3.1. Extinction thresholds across a wide range of dispersal areas*

We estimated the extinction threshold $\lambda_E(A)$ across a wide range of dispersal areas $4 \leq A \leq 51432$ (**Table 2**). Except the standard case *A*=4, these numerical estimates, to our best knowledge, are new. Bramson et al. (1989) and Durrett and Perkins (1999) assumed square neighborhoods of different sizes, but only in the infinite limit $A \to \infty$.

**Table 2.** The extinction threshold $\lambda_E(A)$ at different dispersal areas *A*, grouped according to the lattice geometry. The errors in the last digit of the estimates are shown in parentheses, indicating the range around $\lambda_E(A)$ within which the estimate is reliable. Outside this range, the system was found to be subcritical or supercritical (see the Appendix for details).

| **Diamond** | | **Square** | | **Hexagon** | | **Circle** | |
|---|---|---|---|---|---|---|---|
| ***A*** | $\boldsymbol{\lambda_E(A)}$ | ***A*** | $\boldsymbol{\lambda_E(A)}$ | ***A*** | $\boldsymbol{\lambda_E(A)}$ | ***A*** | $\boldsymbol{\lambda_E(A)}$ |
| 4 | 1.6487(1) | 8 | 1.44070(2) | 6 | 1.5478(1) | 12 | 1.31510(2) |
| 12 | 1.3151(1) | 24 | 1.19528(2) | 18 | 1.2460(1) | 28 | 1.17376(2) |
| 24 | 1.1910(1) | 48 | 1.11576(2) | 36 | 1.1467(1) | 48 | 1.11717(1) |
| 40 | 1.1312(1) | 80 | 1.07827(2) | 60 | 1.0997(1) | 196 | 1.03989(1) |
| 60 | 1.0967(1) | 120 | 1.05715(3) | 90 | 1.0730(1) | 796 | 1.01294(1) |
| 84 | 1.0749(1) | 168 | 1.04390(5) | 126 | 1.0563(1) | 3208 | 1.00404(1) |
| 112 | 1.0599(1) | 224 | 1.03496(3) | 168 | 1.0449(1) | 12852 | 1.00122(1) |
| 144 | 1.0494(1) | 288 | 1.02861(4) | 216 | 1.0368(1) | 51432 | 1.00036(1) |
| 180 | 1.0414(1) | 360 | 1.02392(2) | | | | |
| | | 440 | 1.02033(2) | | | | |

We examined the shift of the threshold relative to the mean-field scenario (i.e., spatially unlimited dispersal),

$$S(A) = \lambda_E(A) - \lambda_E(\infty) = \lambda_E(A) - 1. \qquad \text{Eq. 4}$$

As the dispersal area increases, the extinction threshold approaches the mean-field value of 1, thus *S*(*A*) vanishes. An important finding is that the data obtained for different shapes of the dispersal area fall roughly on a single master curve. This suggests the existence of a unique *S*(*A*) relationship that is independent of lattice geometry.

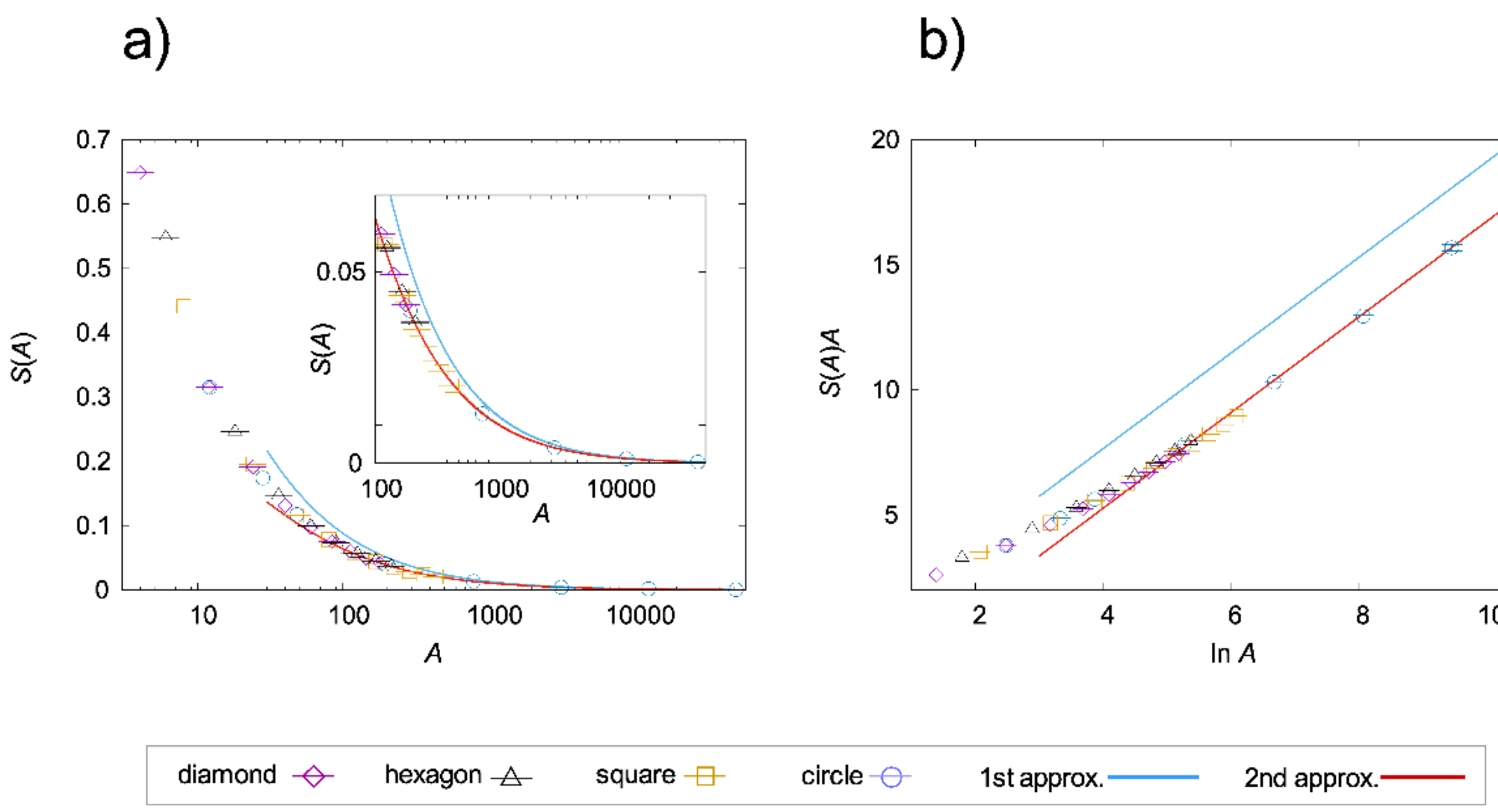

**Figure 4.** Dependence of the shift of the extinction threshold (*S*) on the dispersal area (*A*) in two dimensions. Panel **a)** presents the raw data, while **b)** shows $S(A)A$ in order to check a linear fit. The inset in **a)** magnifies a part of the figure. The symbols show results from the simulations (the estimated shift, and the error of estimation; see Table 2), while the solid lines are theoretical predictions. The first approximation (blue line) provides an estimate for the slope according to **Eq.5**, while the second approximation (red line) yields an estimate for the intercept (**Eq.6**).

To describe the shape of the *S*(*A*) function, first we fitted a theoretically expected curve to the numerical data (**Section 3.2**). We found a good fit within a wide range, from $796 \leq A$ to infinity. We will refer to this range of *A* as the 'large area domain' (LA). Next, we analyzed the data points at lower *A* values (**Section 3.3**), and found that the *S*(*A*) relation could be well approximated by a scaling law over another substantial range ($6 \leq A \leq 224$). We will refer to this range as the 'small area domain' (SA). We examine the results within the LA and SA domains in turn.

### *3.2. The large area (LA) domain*

A key to understanding the system's behaviour in the case of large, finite dispersal areas is to consider its properties in the limit $A \to \infty$. A seminal work has been published in probability theory (Bramson et al., 1989), where the Authors assumed infinite d-dimensional cubic lattices with cubic dispersal neighborhoods, and provided asymptotical formulae for the dependence of $S(A)$ on the generalized dispersal area (i.e., the number of cells in the dispersal neighborhood) in the limit $A \to \infty$. In two dimensions, the following relation has been rigorously proved:

$$S(A) \simeq K(\ln A)/A. \qquad \text{Eq. 5}$$

$K$ denotes a constant, and the relation $\simeq$ expresses that the ratio of the two sides tends to 1 as $A \to \infty$. The right-hand side of **Eq. 5** is thus the leading term of $S(A)$, which tends to zero more slowly than the correction terms in the limit $A \to \infty$. So, it can be considered a first approximation to $S(A)$.

**Eq. 5** shows a scaling law with a scaling exponent -1. It is worth noting that scaling laws also hold in $d \neq 2$ dimensions, but with different exponents (Bramson et al., 1989). **Supplement 2** presents our simulations on a one-dimensional lattice and verifies that the scaling law proposed by Bramson et al. (1989) for $d$=1 under the assumption that $A \to \infty$ holds across a broad range of $A$. Thus, our findings in $d$=1 underscore the generality of scaling behavior at the extinction threshold.

The logarithmic factor in **Eq. 5**, ln $A$, is a specific feature in $d$=2. [It originates from the properties of random walks, which are known to be marginally recurrent in two dimensions (Karlin and Taylor 1975).] Durrett and Perkins (1999) determined the exact value of the constant $K$ in two dimensions and reported it as $K = 6/\pi$. We emphasize that this value was obtained under the assumption of square-shaped neighborhoods. A heuristic derivation of **Eq. 5** was given by Bramson et al. (1989), based on branching random walk. This derivation is valid for any neighborhood of not too anisotropic shape, such as those considered in our work. In addition, Ovaskainen et al. (2020) found **Eq. 5** to hold also for a model that can be considered as a continuous-space analogue of the two-dimensional contact process. These studies motivated us to investigate the scaling relations further. First, it remained unknown whether the value of $K$ is the same across different lattice geometries, beyond the square-shaped neighborhood considered by Durrett and Perkins (1999). Second, **Eq. 5** provided only the leading term to $S(A)$ in the infinite limit. We hypothesized that additional corrections would be necessary for finite values of $A$.

Our results confirm that **Eq. 5** accurately predicts the slope of the $S(A)A$ line (**Figure 4.b** within). As a novel finding, we demonstrate that the slope, $K = 6/\pi$ is unaffected by the underlying lattice geometry. Another new result is an estimate of the line's intercept, $b = -2.4(1)$. Based on this, we propose an improved (second) approximation,

$$S(A) = K\frac{\ln(A/A_0)}{A}, \qquad \text{Eq. 6}$$

where $A_0$ is related to $b$ through $b = -K \cdot \ln A_0$. Its value is approximately $A_0 = 3.5(2)$. Note that $A_0$ does not affect the dominant term (**Eq. 5**) proposed by Bramson et al. (1989), it merely suggests a subdominant term undetermined so far. Introducing the effective dispersal area, defined as $A_e = \frac{\ln(A/A_0)}{A} = \frac{\ln A - 1.25}{A}$, we arrive at a simple scaling law,

$$S(A) = K{A_e}^{-1} \qquad \text{Eq. 7}$$

**Figure 4.a** shows that the second approximation (red curve) successfully predicts the extinction threshold across a broad range of $A$ values (cca. $A \geq 796$). We refer to this range as the large area (LA) domain. At lower $A$ values, however, the data points significantly deviate even from the second approximation. This means that there must be further correction terms beyond the subdominant one, which are suppressed at higher values of *A*, but become relevant at low *A*. To obtain a more complete picture of *S*(*A*), we analyzed the low-*A* data points as well.

### *3.3. The small area (SA) domain*

This domain ranges from *A*=6 (hexagon) to 224 (square). A log-log plot of the numerical data (**Figure 5**) suggests the following scaling relation.

$$S(A) \propto A^{-\mu}. \quad \text{Eq. 8}$$

The symbol ∝ denotes proportionality. The data point *A* = 4 was omitted from the regression analysis due to its marked deviation from the overall trend. This deviation is noteworthy, as the majority of contact process studies on lattices have relied on the four-cell neighborhood, with only a few exceptions (e.g., Bramson et al., 1989; Durrett and Perkins 1999). This finding suggests that the four-cell neighborhood may significantly distort spatial interactions.

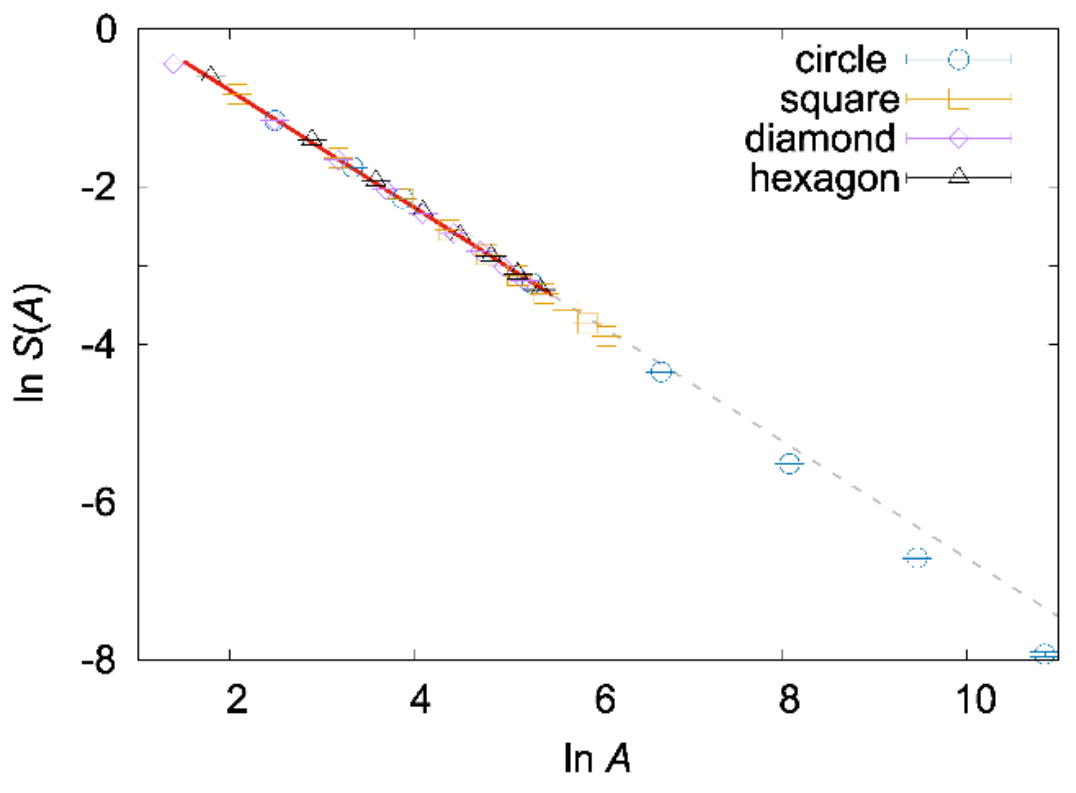

**Figure 5.** Scaling behavior in the small area (SA) domain. The symbols show results from numerical simulations (the estimated shift, and the error of estimation; see Table 2). The red line shows a best-fit linear regression to data obtained across different lattice geometries. The fitting was performed over the range 6 ≤ *A* ≤ 224. The dashed grey line presents an extrapolation, showing the deviation of **Eq. 8** can be adequately fitted to the data with a slope $\mu = 0.74(1)$. We have to note, however, that all data points deviate to some extent from the asymptotic law, with the largest deviation at $A = 4$, corresponding to the smallest neighborhood, where the discreteness of the underlying lattice is most pronounced. Within the SA domain ($6 \leq A \leq 224$), the variation of the local slopes is virtually negligible; therefore, *S*(*A*) can be approximated by the scaling law in **Eq. 8**, with an exponent $\mu = 0.74(1)$.

## 4. Discussion

### *4.1. The extinction threshold*

Accurate estimation of the extinction threshold is of paramount importance with regard to the protection of endangered species and the eradication of invasive weeds. Previous investigations on the contact process have shown that extinction in dispersal-limited populations (i.e., in the case of finite *A*) is analogous with a critical phase transition (from the surviving to the extinct phase); thus even a small decrease in the demographic rate $\lambda$ can lead to a large drop in the population density near the threshold (**Figure 3**). The near-threshold behavior has been studied thoroughly in statistical physics (e.g., Marro and Dickman 1999), and ecology (Oborny et al., 2005), but the value of the threshold has received less attention so far. Most studies in two dimensions have used the minimal dispersal area *A*=4 in, for which the threshold was estimated $\lambda_E(A = 4) = 1.6488(1)$ (Grassberger 1989). The sparsity of research on the general form of $\lambda_E(A)$ can probably be explained by the fact that the threshold value is non-universal, i.e., it depends on the details of the model, such as the lattice geometry and the neighborhood size. Our study focused specifically on these properties.

We considered two kinds of regular lattices (square and hexagonal) and varied the shape of the neighborhood (diamond, square, circle, and hexagon). In each case, we evaluated the shift of the extinction threshold relative to the case of unlimited dispersal. We found the *S*(*A*) relationship independent of the lattice geometry across a very broad range, from *A*=6 to 51,432. (Only the smallest area, *A*=4, was an outlier.) We did not find any *S*(*A*) function generally applicable for the whole range of *A*, but we could highlight two domains (sub-ranges) of *A* within which *S*(*A*) could be approximated with reasonable accuracy (**Table 3**).

**Table 3.** Summary of the results: shift of the extinction threshold in the small-area (SA) and large-area (LA) domains. First, we summarize the general forms (**Eq. 8** for SA, and **Eqs. 6** for LA), and then substitute the estimated constants into the expressions. The value $K = 6/\pi \cong 1.9$ is from Durrett and Perkins (1999; see **Section 3.2**), while $A_0 \cong 3.5$ and $\mu \cong 0.74$ are estimated from our simulations (**Sections 3.2** and **3.3**, respectively).

| Domain | Range of A (# of cells) | Shift of the extinction threshold | |
|---|---|---|---|
| | | **General** | **With the constants** |
| SA | 6 – 224 | $S(A) \propto A^{-\mu}$ | $S(A) \propto A^{-0.74}$ |
| LA | 796 – ∞ | $S(A) = K\frac{\ln(A/A_0)}{A}$ | $S(A) = 1.9\frac{\ln(A/3.5)}{A}$ |

In general, the extinction threshold $\lambda_E(A)$ increases with the decrease of dispersal area *A*, that is, higher *c/e* ratio is needed for survival. This is caused solely by the increase of intraspecific competition due to local aggregation (**Figure 2**).

It is worth discussing the scaling law describing the dependence of the threshold on *A* for large *A* within a broader theoretical context. The extinction threshold is a critical point of the model, at which the spatial and temporal correlation lengths diverge to infinity. In other words, there is no finite characteristic length and time scale appearing in the system's observables. This is the origin of the scaling laws expressed as power laws rather than exponential functions. In this context, the observed power-law shift of the threshold is a natural consequence of the criticality of the system. There is no characteristic dispersal area (apart from the "microscopic" one, i.e, the area of the cell), above which the shift would become negligible. This has significant implications for ecolution from a theoretical point of view. If a characteristic dispersal area existed above which no further fitness gain occurred, then selection would favor dispersal areas not exceeding this characteristic value. According to our findings, this is not the case: there is a slowly decreasing but steady evolutionary pressure favoring increased dispersal areas, a phenomenon deeply rooted in criticality. Conversely, opposing selection pressures may arise due to the costs associated with dispersal (see Section 4.3).

### *4.2. Applicability to plant populations and metapopulations*

To assess the relevance of the results summarized in **Table 3**, we examine the plausible range of *A* in natural populations and metapopulations. To maintain generality across various lattice geometries, we defined the dispersal area *A* as the number of neighboring cells accessible from a focal cell within one dispersal step (**Figure 1**). Converting this *A* into real-world area units (e.g., square meters) requires specifying the lattice constant. We present some examples of lattice constants and discuss the plausible range of *A*.

#### 4.2.1. The large area domain

Let us take an example of a plant population, in which each cell represents a site that can host a single individual. Let us assume that the maximal size of an individual is 10 cm x 10 cm. In this case, the lower limit of the LA domain ($A$=796) corresponds to cca. 1.59 m dispersal distance (assuming circular dispersal). According to **Section 3.2**, the effective dispersal area is $A_e$=147 cells. **Eq. 7** states that the shift of the threshold is inversely proportional to the effective dispersal area. Therefore, doubling this area would halve the shift, bringing the population closer to unlimited dispersal. 1.59 m radius falls into the realistic range of seed dispersal (c.f. (Jenkins et al., 2007), Lososová et al., 2023). Many wind-dispersed and animal-dispersed seeds have much larger radii (Munoz et al., 2013; Bullock et al., 2017) and thus belong to the acceptable domain. It must be mentioned, however, that species with fat-tailed dispersal kernels generally fall outside the scope of our model, because no characteristic area of dispersal can be defined. In the case of populations, each cell is vacant or occupied by a single individual. In a site-occupancy model of a metapopulation, however, a cell can represent a large area (even a hectare or more). In this case, the required minimum area, $A$=796, can hardly be reached or surpassed, except for very good dispersers.

#### 4.2.2. The small area domain

Taking the previous example about a plant population with a 10 cm x 10 cm unit area, the corresponding dispersal distance would range from 10 to 70 cm in the SA domain. In the former case, the distance is extremely small: the plant individual can reproduce only in adjacent cells. This can be realistic in some cases of vegetative reproduction (for example, when new bulbs are produced adjacent to the parent). But the SA domain is unlikely to apply to seed dispersal, at least in the case of small-bodied plants. In contrast, let us consider a large-bodied plant (e.g., a tree) with a 5 m x 5 m spatial unit. In this case, the dispersal distance should be $\leq$ 35 m. This condition is satisfied by many tree species (cf. Clark et al., 1999; Munoz et al., 2013; Bullock et al., 2017). Additionally, in metapopulation models with large spatial units (e.g., a hectare), the SA domain is generally more realistic than the LA domain.

In summary, the scaling law in the SA domain (**Table 3**) is likely applicable to metapopulations and populations of large-bodied plants, whereas the law in the LA domain may be more appropriate for smaller-sized plants. There is an intermediate zone between the SA and the LA domains within which our simulations do not suggest any simple $S(A)$ relationship. Exploring the entire range of $A$ is an exciting matter for future research.

### *4.3. Perspective on the evolution of dispersal*

Understanding the evolution of dispersal areas necessitates the consideration of some further questions. 1) Is natural selection sufficiently strong to drive an increase in $A$ from its current value? 2) What is the comparative advantage of increasing $A$ relative to increasing $\lambda$? 3) What are the main trade-offs in biomass allocation that constrain this evolutionary process?

The answers, in general, depend on the current state ($A$ and $\lambda$), the costs of increasing $A$ vs. $\lambda$ in terms of resources, and the benefit in terms of fitness. The potential changes are generally constrained by the total amount of resource $R$ and by allocation trade-offs, which restrict the set of feasible ($A$, $\lambda$) pairs. Altogether, this is a complex optimization problem on the $w(A, \lambda)$ fitness landscape, in which the parameters depend on the species and environment. Our results cannot fully answer the questions, however, they offer some useful clues.

*1) The strength of selection.* Our results suggest that selection weakens strictly monotonously with $A$ in the vicinity of extinction. This is indicated within the whole range of $A$ by the strictly monotonous decrease of the derivative $S(A)$, approaching 0 (**Figure 4a**). To show this decline more directly, we have plotted $A$ against $\lambda_E = S(A) + 1$ using linearly scaled axes (dots and bold line in **Figure 6**). Let us take the example of $A$=168, which is quite close to the upper limit of the SA domain. Dispersal limitation can be characterized by comparing the actual $\lambda_E(A)$ to the unlimited $\lambda_E(\infty) = 1$, i.e., by the vertical distance between the dot and the dashed line. This is $\lambda_E(168) - \lambda_E(\infty) = 0.044$ (see the data in **Table 3**; here rounded to three digits). Increasing the dispersal area to $A$=224 (indicated by an arrow in **Figure 6**), the extinction threshold shifts by $\lambda_E(168) - \lambda_E(224) = 0.009$. This is a considerable change. Taking the example of the above-mentioned tree species with a 5 m x 5 m unit area, the result suggests that increasing the dispersal distance from 30 m to 35 m would shift the extinction threshold by roughly 1%. Thus, assuming a constant local extinction rate *e*, the colonization rate *c* can be roughly 1% lower for successful population survival. This kind of improvement is even larger at smaller values of *A*.

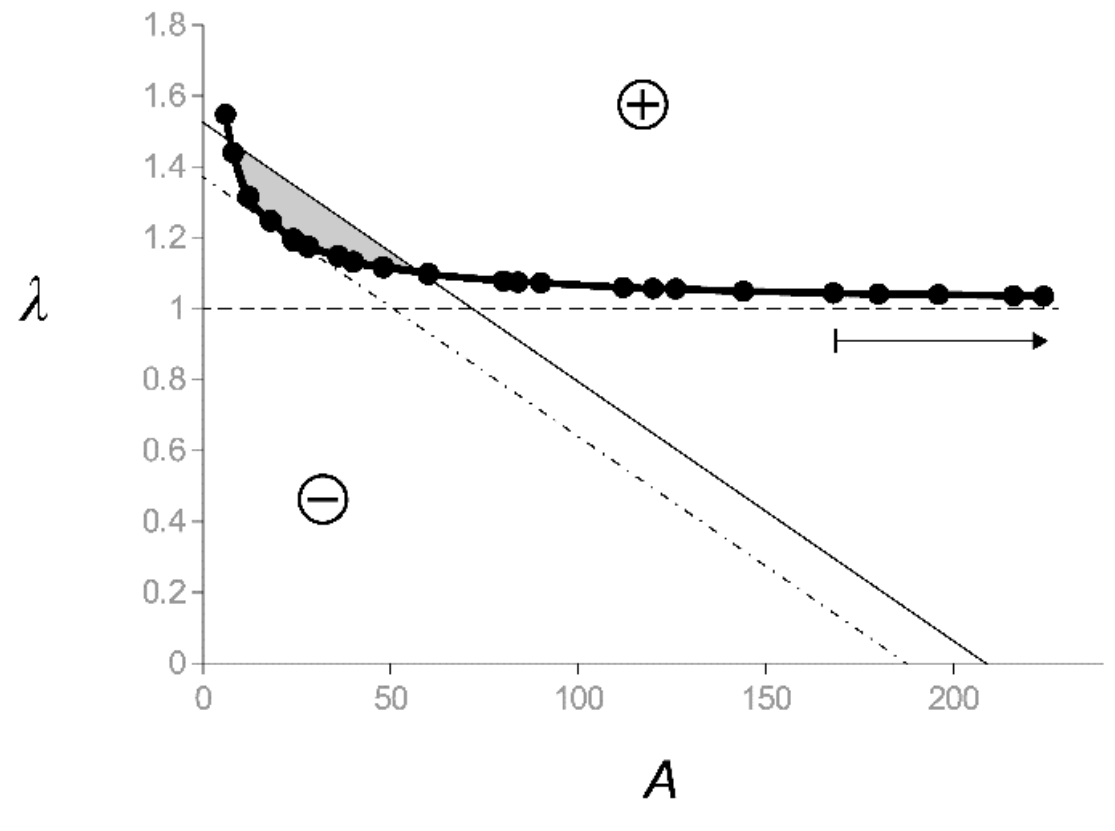

**Figure 6.** Relationship between the dispersal area $A$ and the demographic parameter $\lambda=c/e$ within the SA domain $(6 \leq A \leq 224)$. The dots and the bold line show our results on the extinction threshold $\lambda_E(A)$ separating positive from negative population growth (see the signs in circles). The horizontal dashed line shows the case of unlimited dispersal, $\lambda_E(\infty) = 1$. The arrow indicates a hypothetical increase in A. The thin solid line and the dashed-dotted line show examples of a trade-off between $A$ and $\lambda$. The dashed-dotted line is tangent to the $\lambda_E(A)$ curve (see text for details).

Conversely, the LA domain is characterized by weak dispersal limitation and weak selection for increasing *A*. For higher performance, it is presumably more advantageous to increase the probability of colonization (*c*) or survival in the colonized sites (1-*e*).

*2) Trade-offs and comparative advantage.* Let's begin by distinguishing between spatial versus non-spatial events in our model. Dispersal was a spatial event, as it always linked two sites (a seed donor and a recipient). Colonization, on the other hand, was non-spatial, as it consisted of local events either within the donor or the recipient cell (**Figure 7**). Survival of the reproductive individuals was also local. In the broader literature, models of dispersal kernels typically focus on step 1 (e.g., Bullock et al., 2017), while non-spatial demographic models, such as projection matrix models (Groenendael et al., 1988; Salguero-Gómez et al., 2015), typically account for events 2-5. It is noteworthy that some authors have proposed considering the spatial pattern of seed dispersal in conjunction with that of germination and juvenile survival (Schupp and Fuentes 1995; Schupp et al., 2010; Rogers et al., 2019), that is, integrating 1, 2, and 3. Bonte et al. (2012) and Travis et al. (2012) published comprehensive reviews exploring the potential trade-offs between dispersal and various life-history events.

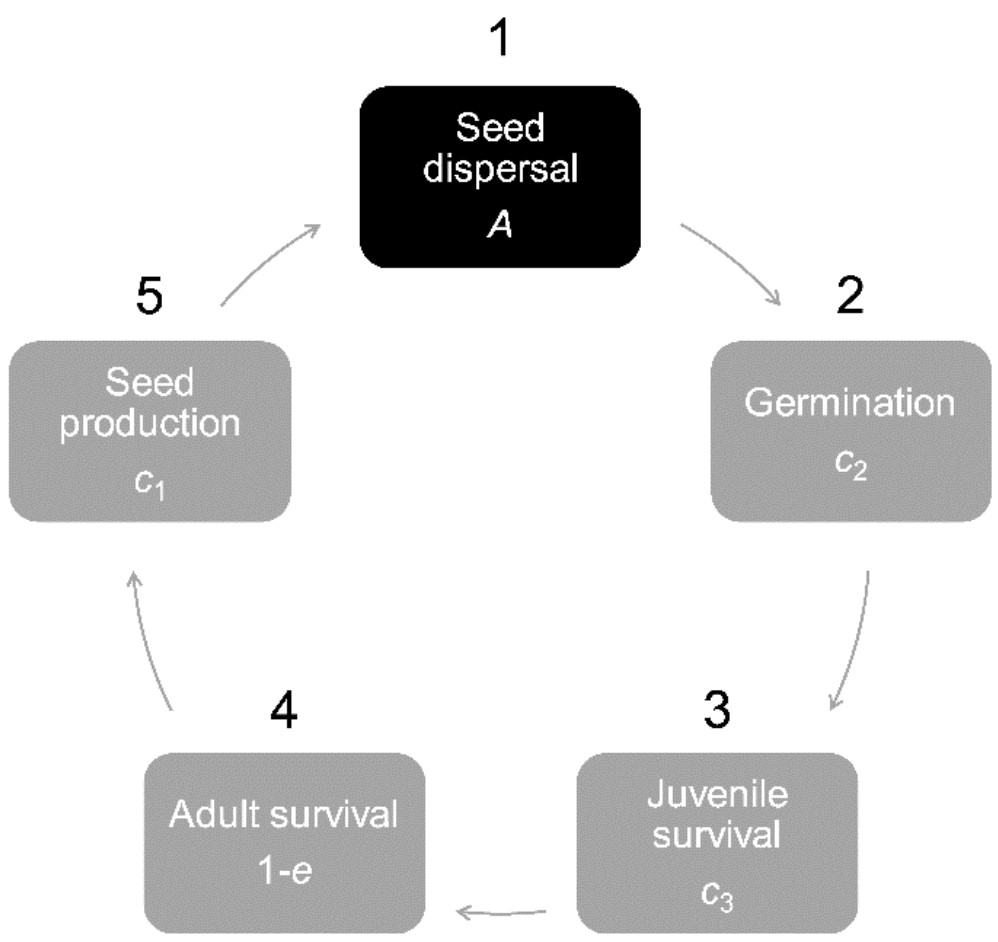

**Figure 7.** Simplified life cycle, subdivided into spatial (black) and non-spatial events (grey). The spatial event links two sites, while the non-spatial events occur locally, within a single site. The corresponding variables are written in italics. The probability of colonization is a function of the probability of its constituent steps, $c(c_1, c_2, c_3)$.

The diagram emphasizes that *c* can be increased before dispersal (increasing seed production), or after dispersal (increasing the probability of germination, or survival of the new individuals up to reproductive age). This is a simplified view, as it does not account for age-dependence or phenotypic plasticity. Nevertheless, even this simple model involves immense complexity due to the interdependencies between the variables. Trade-offs have been shown, for example, between the quality and quantity of seeds (Eriksson and Jakobsson 1999), thus between 2 and 5 in **Figure 7**. Larger seed size may decrease the dispersal area, but increase the probability of germination (Tuthill et al., 2023), linking 1 and 2. Increasing seed production may decrease adult survival (Ehrlén and van Groenendael 1998), connecting 4 with 5. Multiple dependencies are also possible (Ehrlén and van Groenendael 1998; Nathan and Muller-Landau 2000; Weigang and Kisdi 2015). Addressing all of these factors is outside the scope of this paper, we only present an example of a trade-off between *A* and $\lambda$.

For the sake of simplicity, let us assume a linear trade-off,

$$R = aA + b\lambda, \qquad \text{Eq. 16}$$

where a fixed amount of resource *R* can be allocated into increasing *A* or $\lambda$, with efficiencies *a* and *b*, respectively. It follows from **Eq. 16** that changing the amount of resource in the environment shifts the line along the vertical axis without changing its slope (see the solid and dotted lines in **Figure 6**). In the case of the solid line, the set of viable (*A*, $\lambda$) combinations is marked by grey shading. These points are above the extinction threshold (solid bold curve) and are affordable in the sense that they are below the trade-off line (solid thin line). Along a trade-off line, *R* is fully utilized; thus, the (*A*, $\lambda$) combinations above the line are not affordable under the current resource limitation. The dotted line is specific in the sense that it is tangent to the extinction curve with the given slope. Thus, no (*A*, $\lambda$) combination is viable below this line. The shape of the bold curve, obtained from our simulations, suggests that the tangent line flattens rapidly with the increase of $A$, and thus the range of viable ($A$, $\lambda$) combinations becomes primarily determined by $\lambda$.

We propose the following answers to the questions posed at the beginning of this section. The strength of selection upon $A$ declines rapidly with the increase of $A$. At the beginning of the SA domain, the population's viability depends on the interplay between $A$ and $\lambda$, including their trade-off. Towards the end of the SA domain (and beyond), viability becomes dominated by $\lambda$. It is important to note, however, that our statement is restricted to the vicinity of extinction [i.e., the zero isoline of the fitness landscape $w(A, \lambda)$], and is not applicable to the whole fitness landscape.

## Summary and outlook

The extinction threshold was always higher in cases of limited dispersal (within area $A$) than in the case of unlimited dispersal; that is, survival was restricted to a smaller parameter range. The shift in the threshold, $S(A)$, was greatest at the lowest value of $A$=4 (i.e., in the classical contact process), and then decreased monotonously. Notably, $S(A)$ could be described by a universal curve, which was independent of the lattice geometry (except for $A$=4, which was an outlier). The simulations revealed that $S(A)$ could be described by scaling laws in two distinct, rather broad ranges of $A$. To fully understand the role of $A$ in population extinction, it would be essential to determine the $S(A)$ function across the whole range of $A$.

The shape of the $S(A)$ curve suggests potential evolutionary implications. Selection for increasing $A$ seems to be considerable within the whole SA domain but appears negligible within the LA domain. However, it is important to note that this statement is restricted to homogeneous environments, where the only disadvantage of small $A$ is clumping (**Figure 2**), leading to enhanced intraspecific competition. In models with two or more species, escaping from interspecific competition is also an important factor (cf. the competition-colonization trade-off; Ehrlén and van Groenendael 1998; Bolker and Pacala 1999; Higgins and Cain 2002). A site may be unsuitable due to occupancy by another species, or because of heterogeneity in the environmental background. In these cases, moving from one suitable patch to another becomes a crucial factor (see, for example, Hanski 1994; King and With 2002; Ronce 2007; Baguette et al., 2013; Bocedi et al., 2014; Finand et al., 2024). In general, dispersal is subject to multiple selection pressures (see Ronce 2007; Duputié and Massol 2013; and Saastamionen et al. 2018 for comprehensive reviews). Our simulations suggest that avoiding clumping is a strong selection factor in the SA domain, but weakens at higher *A*. In the LA domain, the evolution of *A* is likely influenced by additional factors that partition space into suitable versus unsuitable terrain.

Our study assumed that seeds were dispersed uniformly within *A*; i.e., the dispersal kernel was a step function. Several papers emphasize the diversity of dispersal kernels (Schupp and Fuentes 1995; Bullock et al., 2017; Rogers et al., 2019). Examining the influence of alternative dispersal kernels would be essential to extend the scope of our findings. Another important factor is the accumulation of seed bank. Our model assumed immediate germination and establishment. Incorporating a seed bank would introduce a time delay into the model, and thus relax the present assumption of purely contact interactions within the neighborhood *A*. Investigating the effects of such a delay, caused by the addition of a second (soil) layer in the lattice, would be an exciting task for future research. A further extension of the present scope would be to link our parameters to plant traits (seed size, plant height, etc.; c.f. Ehrlén and van Groenendael 1998) according to the approach illustrated in **Figure 6**.

## Acknowledgements

We are grateful to Máté Gulyás and Anna-Mária Csergő for inspiring discussions about the subject. Funding: This work was supported by the National Research, Development and Innovation Office NKFIH [grant numbers K146736, K143622].

## APPENDIX

### *Estimation of the extinction threshold*

The extinction threshold was estimated from some dynamic properties of the spreading process from a single occupied site in the center of the lattice (at $t$=0). The simulations ran in $10^4$–$10^6$ independent repetitions (trials) in the case of every parameter combination ($A$, $\lambda$). To preclude finite-size effects, the simulations were terminated as soon as any run in the set of trials reached the edge of the lattice. The time-dependence of three quantities was recorded:

1) $P(t)$, the proportion of surviving trials,

2) $\langle N(t)\rangle$, the mean number of occupied sites, and

3) $\langle R^2(t)\rangle$, the mean squared displacement.

The bracket $\langle . \rangle$ denotes averaging over all the trials. $R^2$ is the average squared distance of the occupied sites from the origin,

$$\langle R^2(t)\rangle = \frac{1}{\langle N(t)\rangle}\langle \sum_{i=1}^{M} r_i^2\, \sigma_i\rangle \qquad \text{Eq. A1}$$

where $i$ is the site's label ranging from 1 to the total number of sites ($M$), $r_i$ is the Euclidean distance of site $i$ from the origin, and $\sigma_i$=1 if the site is occupied, and 0 if it is vacant.

Due to the critical phase transition at the extinction threshold, these quantities obey scaling laws in the limit of $t \to \infty$, as revealed by Grassberger and de la Torre (1979):

$$P(t) \propto t^{-\delta},$$

$$\langle N(t)\rangle \propto t^{\eta}, \qquad \text{Eqs. A2}$$

$$\langle R^2(t)\rangle \propto t^{\zeta}.$$

The values of the scaling exponents ($\delta$, $\eta$, and $\zeta$) are universal, i.e., their values depend only on the dimensionality (now $d$=2), and are independent of the lattice geometry and neighborhood size for all contact processes (and more broadly, within the directed percolation universality class; Broadbent and Hammersley 1957). The exponents in two dimensions are $\delta$=0.451(3), $\eta$=0.229(3), and $\zeta$=1.133(2), where the error in the last digit is shown in parentheses (Grassberger and Zhang 1996; see other estimations in Dickman 1999; Arashiro and Tomé 2007).

According to the scaling relations (**Eqs. A2**), $P(t)$, $\langle N(t)\rangle$, and $\langle R^2(t)\rangle$ tend to straight lines on log-log plots as the extinction threshold is approached, i.e., $|\lambda - \lambda_E(A)| \to 0$. We used this relationship for estimating the threshold, observing the curvatures of $P(t)$, $\langle N(t)\rangle$, and $\langle R^2(t)\rangle$ when increasing $\lambda$ in small steps (**Figure A1**). The figure illustrates the characteristic behavior of a spreading (meta)population. The survival probability $P(t)$ tends to a constant value above the extinction threshold [ $\lambda > \lambda_E(A)$], while it declines exponentially below the threshold [ $\lambda < \lambda_E(A)$]. The mean number of occupied sites $N(t)$ increases proportionally to $t^d$ at large t values above the threshold and declines exponentially below. The mean squared displacement $\langle R^2(t)\rangle$ follows a quadratic increase above the threshold and tends to a linear increase below (Marro and Dickman 1999). Importantly, all three observables approach straight lines as the distance from the threshold vanishes, $|\lambda - \lambda_E(A)| \to 0$. Conversely, the larger the distance from the threshold, the sooner the deviation from the straight line starts.

We estimated the extinction threshold from $\langle N(t)\rangle$. We preferred this observable to $P(t)$ or $\langle R^2(t)\rangle$ because its deviations from a straight line were easier to discern. The threshold was estimated by successive approximation, narrowing the difference between its upper and lower bound gradually. Since the fluctuations increased towards the threshold (c.f. Harris 1974), we increased the number of trials from $10^4$ to $10^6$, and the lattice size from 500x500 to 2000x2000 cells near the threshold. This procedure allowed us to estimate the thresholds up to four digits (**Table 2** in the main text).

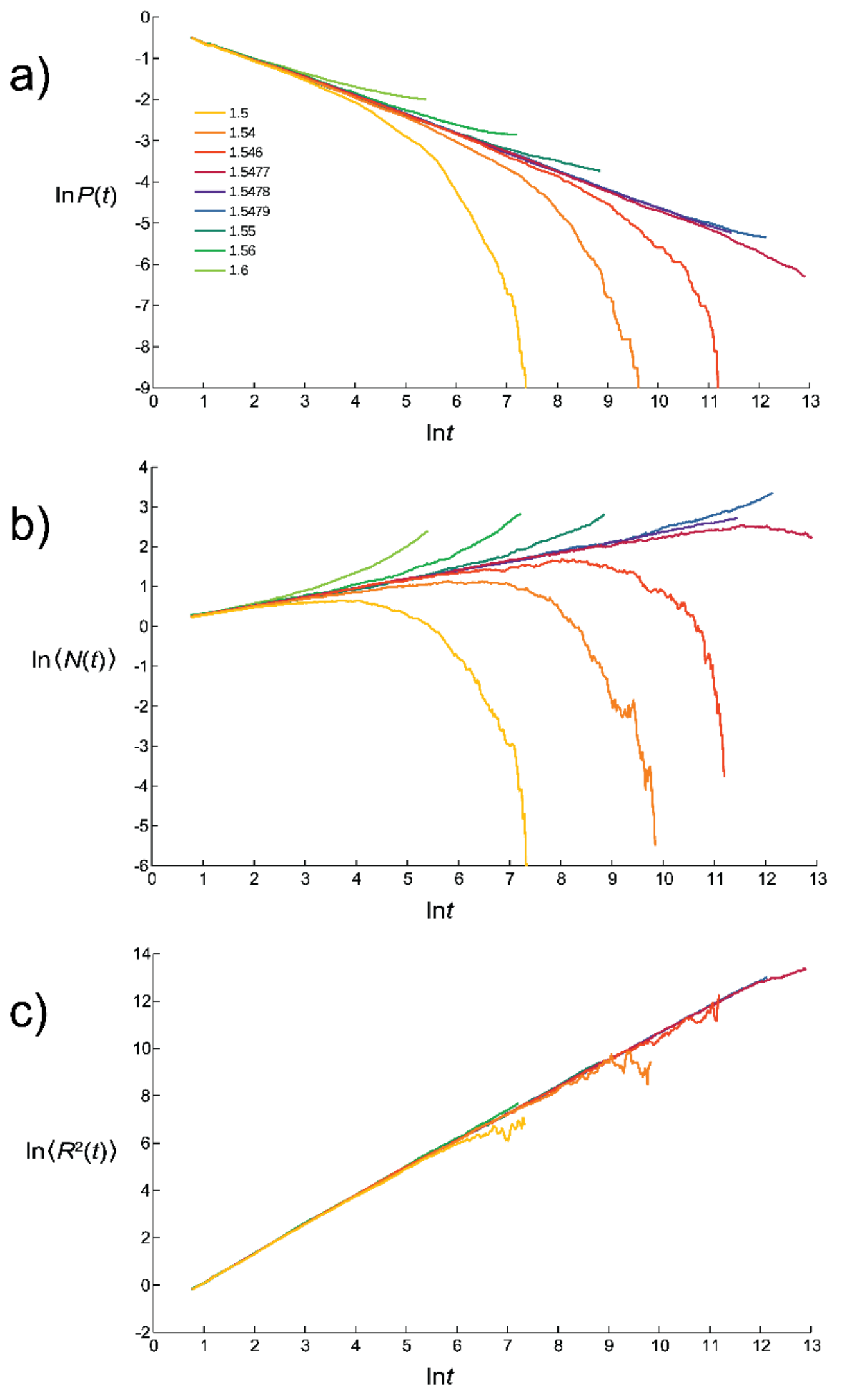

**Figure A1.** Expansion of a metapopulation starting from a single occupied site. **a)** The proportion of surviving trials $P(t)$, **b)** the mean number of occupied sites $\langle N(t)\rangle$, and **c)** the mean squared displacement $\langle R^2(t)\rangle$. The figure shows an example in a hexagonal lattice with a dispersal area *A*=6. The demographic parameter $\lambda$ increased from 1.5 to 1.6 as shown by the color code.

## References in the Appendix